\documentclass[conference]{IEEEtran}
\usepackage{mdframed}
\usepackage{float}
\usepackage{graphicx}
\usepackage{tikz}
\usetikzlibrary{shadows}
\usepackage{xurl}
\usepackage{url}
\usepackage{hyperref}
\usepackage{titlesec}
\titlespacing*{\section}{0pt}{8pt plus 2pt minus 2pt}{4pt}
\titlespacing*{\subsection}{0pt}{6pt plus 2pt minus 2pt}{3pt}

\newmdenv[
  linewidth=0.5pt,
  roundcorner=2pt,
  innertopmargin=4pt,
  innerbottommargin=4pt,
  innerleftmargin=6pt,
  innerrightmargin=6pt
]{answerbox}

\begin{document}

\title{All Smoke, No Alarm: Oracle Signals in Agent-Authored Test Code}

\author{
\IEEEauthorblockN{Dipayan Banik}
\IEEEauthorblockA{\textit{Danovo Energy Solutions} \\
Raleigh, North Carolina, USA \\
dbanik@danovoenergy.com}
\and
\IEEEauthorblockN{Kowshik Chowdhury}
\IEEEauthorblockA{\textit{Kennesaw State University} \\
Kennesaw, Georgia, USA \\
kchowdh1@students.kennesaw.edu}
\and
\IEEEauthorblockN{Shazibul Islam Shamim}
\IEEEauthorblockA{\textit{Kennesaw State University} \\
Kennesaw, Georgia, USA \\
mshamim@kennesaw.edu}
}


 % \includegraphics[width=\textwidth]{sampleteaser}
 % \caption{Seattle Mariners at Spring Training, 2010.}
 % \Description{Enjoying the baseball game from the third-base
 %% seats. Ichiro Suzuki preparing to bat.}
  %\label{fig:teaser}

%%\received[revised]{12 March 2009}

\maketitle

\begin{abstract}
Software practitioners increasingly use AI coding agents that generate test code alongside production code in open‑source pull requests (PRs). Recent studies report more than 932,000 agent-authored PRs across more than 116,000 repositories, yet whether their test files contain meaningful verification logic remains underexplored. Test files lacking explicit assertions execute code without verifying behavior, so quality gates based on test-file presence overestimate verification strength. The goal of this paper is to help practitioners assess the verification strength of agent-authored patches by characterizing oracle signals and their link to merge outcomes and review effort. We conduct an empirical study of 86,156 test-file patches from 33,596 agent-authored PRs across 2,807 GitHub repositories produced by five coding agents: OpenAI Codex, GitHub Copilot, Devin, Cursor, and Claude Code. A qualitative analysis of 384 stratified patches informs a syntactic taxonomy of eight oracle signal categories. Applied at scale, 80.2\% of test patches contain weak or no explicit oracle signals. While raw merge rates are lower for strong-oracle PRs, a regression analysis adjusting for agent, PR size, repository popularity, task type, and language shows strong oracles significantly improve merge likelihood (OR = 1.28, \(p < 0.001\)). Our findings suggest that test file counts substantially overestimate verification strength and that practitioners can adopt oracle-aware quality checks to more accurately evaluate agent-authored contributions.

\end{abstract}


\section{Introduction}

The landscape of software development is shifting from AI-assisted code completion to autonomous code authorship. Interactive assistants such as early GitHub Copilot offered line-level suggestions that developers could accept or ignore, preserving human control over commits \cite{li2025rise}. A new generation of autonomous agents, including OpenAI Codex, Devin, Cursor, and Claude Code operates at the repository level: given an issue or prompt, they plan implementations, write production code, generate tests, open pull requests, and in some cases iterate on reviewer feedback without human intervention \cite{reddington2025copilot,watanabe2025agentic}. Industry projections estimate that 75\% of enterprise software engineers will use AI code assistants by 2028 \cite{gartner2024ai}. This transition represents not merely a productivity gain but a fundamental change in who authors the verification logic that guards against regression. 

As agents take on this authorship, the quality of the test code they produce becomes a critical concern. In a typical pull request, a developer writes a test that calls a function and then asserts that the returned value matches an expected result. This assertion is the test oracle, which determines whether the test passes or fails \cite{barr2015oracle}. Without it, a test executes code but never checks whether the output is correct. A pull request that includes test files appears well-verified to a reviewer scanning a file list: CI pipelines report coverage, dashboards turn green, and the contribution looks complete. However, practitioners are increasingly reporting a gap between this appearance and the underlying verification strength. Developers describe AI-generated tests that ``confirm your code does what it already does"— a phenomenon called ``test theater" \cite{houston2025theater}. Common patterns include assertion-free functions that execute code paths but never verify the output \cite{flores2026audit}, and test suites that mock the module under test itself, passing by construction rather than by correctness \cite{barzeev2026worse}.

Research on LLM-generated test oracles reinforces this concern. Controlled experiments show that LLM-generated assertions frequently encode actual program behavior rather than expected behavior, turning bugs into passing tests \cite{konstantinou2024llmoracles}. Fine-tuned assertion generators detect only a fraction of known faults even at 59.5\% exact-match precision \cite{fraser2025assert5}, and mutation-guided feedback improves oracle quality but requires curated prompts unavailable in autonomous agent workflows \cite{khomh2024mutap}. These findings expose a gap unaddressed by existing metrics: code coverage correlates only weakly with fault‑detection effectiveness \cite{inozemtseva2014coverage}, with mature test suites showing gaps of up to 51\% between executed and checked code \cite{hossain2023coverage}. A systematic categorization of what agents actually produce in test patches ranging from no oracle signal to strong behavioral assertions would give both CI tooling and human reviewers a vocabulary to distinguish verified contributions from structural scaffolding. Understanding how these categories relate to merge outcomes and review effort would further clarify whether oracle strength carries practical consequences in real-world development workflows, or whether reviewers treat all test-bearing PRs alike regardless of assertion quality.

Existing research has examined oracle quality in controlled settings and test file presence in real-world repositories, but not the oracle content of agent-authored test patches at scale. We address this gap through a large-scale empirical study of 86,156 cumulative test-file patches from 33,596 agent-authored pull requests across 2,807 GitHub repositories from the AIDev dataset \cite{li2026aidev}. We investigate two research questions: 

\begin{itemize}
\item \textit{RQ1: What oracle signal patterns characterize agent-authored test-file patches?}
\item \textit{RQ2: How does oracle signal strength relate to PR merge outcomes and human review effort?}
\end{itemize}

This paper makes three primary contributions:
(1) We propose a syntactic taxonomy of eight oracle signal categories for classifying test-file patches.
(2) We characterize oracle signal distributions across five AI coding agents.
(3) We analyze the relationship between oracle signal strength and PR merge outcomes through a multivariate logistic regression.

\section{Methodology}

We use the AIDev-pop subset of the AIDev dataset \cite{li2026aidev}, which contains 33,596 agent-authored pull requests across 2,807 repositories with at least 100 GitHub stars. We extract file-level patches from the \texttt{pr\_commit\_details} table, which stores the diff for each file modified in each commit within a pull request. Our analysis scripts, classified dataset, and figures are available at \cite{replication2026}.

\subsection{RQ1: Oracle Signal Classification}
\textbf{Test File Identification:} We identify test files using two criteria applied jointly. First, the file path must match test-related directory or filename patterns, including \texttt{test/}, \texttt{tests/}, \texttt{\_\_tests\_\_/}, \texttt{testing/}, or filenames containing \texttt{\_test.}, \texttt{.test.}, or \texttt{.spec.} Second, the file extension must belong to a set of 30 source code extensions spanning Python, TypeScript, JavaScript, Go, Java, C\#, Rust, Ruby, and others. We exclude non-code artifacts (fixtures, snapshots, documentation) that appear in test directories but contain no executable test logic. This filtering yields 103,976 test code patches from the total of 711,923 file patches.
\textbf{Cumulative Patch Construction:} A single test file may appear in multiple commits within the same pull request. To avoid double-counting and capture the full oracle signal introduced by the PR, we concatenate all patches for each unique (PR, filename) pair in commit order. This produces 86,156 cumulative patches for our taxonomy analysis. \textbf{Oracle Signal Taxonomy:} For each cumulative patch, we extract added lines and search for assertion patterns from widely used testing frameworks (e.g., JUnit, pytest, Jest, Go testing, RSpec). Through iterative open coding, we find that oracle signals fall along a spectrum -- from no assertion to checks that code merely executes, to checks of specific output values or behavior -- which we group into eight categories in Table~\ref{tab:oracle_signal_taxonomy}. Weak categories (W1--W5) confirm execution without verifying output correctness, mirroring the documented gap between executed and output checked code \cite{inozemtseva2014coverage, hossain2023coverage}; strong categories (S1--S3) compare output to expected values, verify error handling or types, or combine multiple assertion types. To assess reliability, two authors independently labeled a stratified random sample of 384 patches (95\% confidence, 5\% margin); we resolved disagreements by discussion and reported agreement with Cohen's Kappa \cite{landis1977kappa}.

\begin{table}[t]
\centering
\caption{Oracle signal taxonomy categories.}
\label{tab:oracle_signal_taxonomy}
\small
\setlength{\tabcolsep}{4pt}
\begin{tabular}{@{}p{0.08\linewidth}p{0.12\linewidth}p{0.70\linewidth}@{}}
\hline
Code & Tier & Definition \\ 
\hline
W1 & Weak & No assertion pattern present \\
W2 & Weak & Existence/non-null checks only \\
W3 & Weak & Boolean asserts only (no value compared) \\
W4 & Weak & Mock/call-verification only \\
W5 & Weak & Snapshot match only \\
S1 & Strong & Value equality or comparison \\
S2 & Strong & Error, containment, or type checks \\
S3 & Strong & Two or more distinct strong types \\
\hline
\end{tabular}
\end{table}

\subsection{RQ2: Oracle Signals and PR Outcomes}

\textbf{PR-Level Quality Groups:} We aggregate patch-level classifications to the PR level by assigning each PR the highest oracle-signal category among its test-file patches. We define three quality groups: \emph{all\_weak} (best patch is W1--W5), \emph{has\_strong} (best patch is S1 or S2), and \emph{has\_multi\_strong} (best patch is S3).
\textbf{Outcome Variables:} We collect merge outcome, review effort (formal reviews + inline + discussion comments), PR size (files changed, lines added), repository popularity (star count), and time to close for each PR.
\textbf{Controlled Comparisons:} We compare the three quality groups across all outcome variables. To control for confounding factors, we repeat the comparison within each agent, within each task type, and within PR size buckets (small 1--5 files, medium 6--20, large $>$20). We apply size-bucket controls to both merge rate and review effort to distinguish whether oracle signal strength has an independent association with PR outcomes. We additionally fit a multivariate logistic regression including agent, log-transformed PR size, log-transformed repository stars, task type, and primary language as covariates.

\section{Analysis and Findings}
\subsection{RQ1: Oracle Signals in Test-File Patches}

\begin{figure}[H]
\centering
\includegraphics[width=1.05\linewidth]{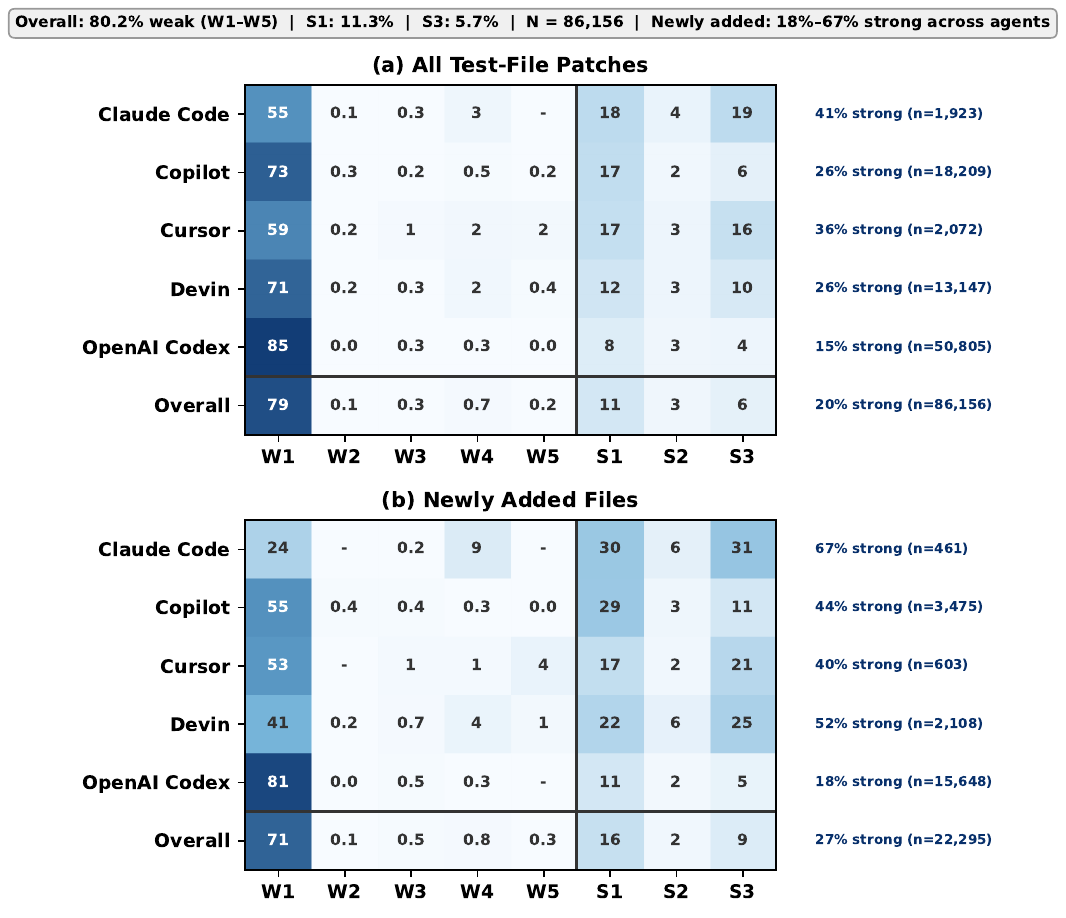}
\caption{Oracle signal distributions and breakdowns for RQ1 analyses.}
\label{fig:rq1}
\end{figure}

Figure~\ref{fig:rq1} presents the oracle signal distribution across all 86,156 cumulative test-file patches and for newly created files separately. As shown in the summary header, 80.2\% of patches contain weak or no explicit oracle signals (W1--W5), with value assertions (S1) and multi-signal strong oracles (S3) accounting for 11.3\% and 5.7\%, respectively. Two authors independently labeled 384 stratified patches (95\% confidence, 5\% margin), reaching Cohen's $\kappa = 0.77$. On the same sample, the classifier matches the human labels for the oracle-signal category in 86.7\% of patches.

Two patterns emerge from the per-agent breakdown. First, agents differ in how often they introduce oracle signals (\(\chi^2 = 2497.3\), \(p < 0.001\)), with Claude Code and Devin producing stronger oracle profiles than Copilot, Cursor, and OpenAI Codex. Second, newly created files show higher strong-oracle rates than modified files (\(\chi^2 = 810.2\), \(p < 0.001\)), ranging from 18\% for OpenAI Codex to 67\% for Claude Code. The newly created-file comparison carries the strongest claim because the full file content is the patch and no prior oracle content exists. \textbf{Key Finding.} \textit{80.2\% of agent-authored test-file patches contain weak or no explicit oracle signals. On newly added test files, strong-oracle rates range from 18\% to 67\% across agents.}

\subsection{RQ2: Oracle Strength, Merge Outcomes, and Review Effort}

\begin{figure*}[t]
\centering
\includegraphics[width=\textwidth]{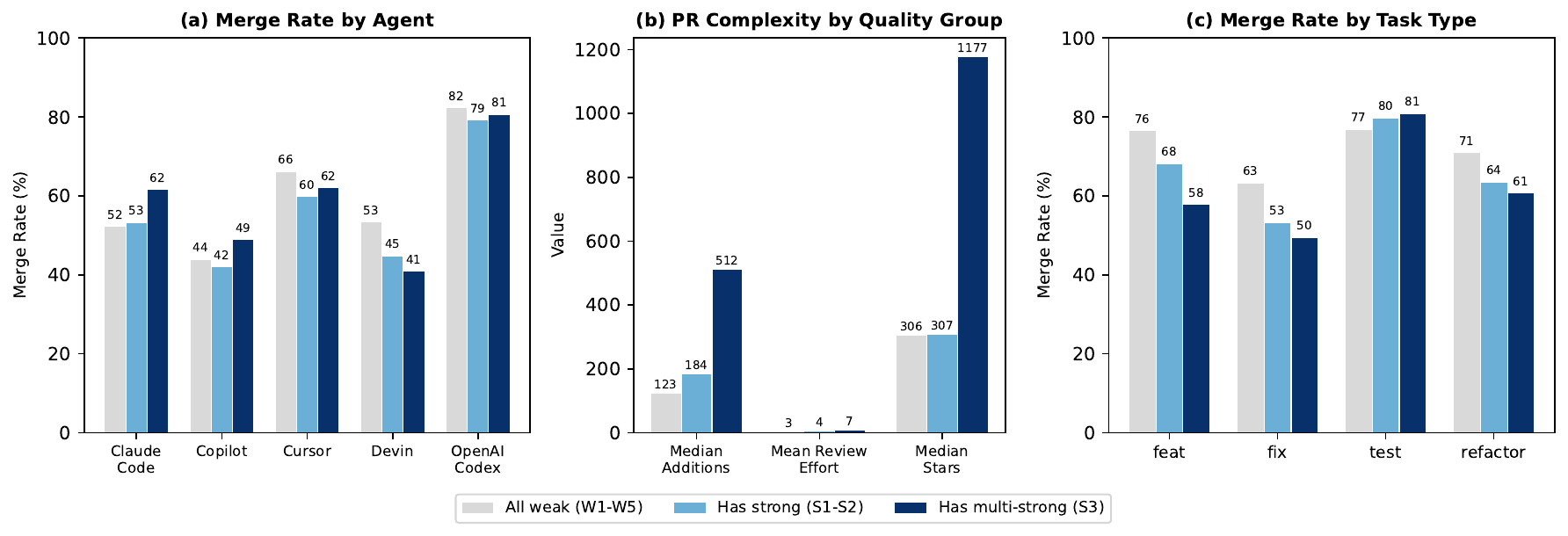}
\caption{Merge outcomes and confounding-factor analyses for RQ2.}
\label{fig:rq2}
\end{figure*}

S3 PRs merge at 59.7\% compared to 72.6\% for weak-oracle PRs, but S3 PRs also contain 4.2$\times$ more code additions (Mann-Whitney test, \(p<0.001\), r = 0.38), attract 2.4$\times$ more review effort (Mann-Whitney test, \(p<0.001\), r = 0.36), and reside in repositories with 3.8$\times$ more stars. This review effort gap persists within PR size buckets, with ratios ranging from 1.7$\times$ for small PRs to 2.5$\times$ for large PRs.

Figure~\ref{fig:rq2} shows that the aggregate pattern weakens within individual agents. Within Claude Code, S3 PRs merge at 62\% compared to 52\% for weak-oracle PRs; Copilot shows the same direction at 49\% versus 44\%. For test-type PRs, S3 PRs merge at 81\% compared to 77\%.  A multivariate logistic regression adjusting for agent, PR size, repository popularity, task type, and primary language confirms that S3 oracles associate with significantly higher merge rates after adjustment (OR = 1.28, \(p<0.001\)).
\textbf{Key Finding.} \textit{Stronger oracle signals are associated with more complex, heavily reviewed pull requests. After adjusting for agent, PR size, repository popularity, task type, and language, S3 oracles significantly increase the likelihood of merge.}

\section{Discussion}

Our findings highlight a meaningful gap between test-file presence and oracle quality in agent-authored pull requests. This pattern may reflect an optimization mismatch: coding agents are trained on code completion objectives that reward structural plausibility, that is, creating test files, importing frameworks, and defining function signatures rather than behavioral reasoning about expected program output. Generating a correct oracle requires an understanding of intended semantics, a capability that current agent architectures do not explicitly optimize for. The variation across agents (18\% to 67\% strong-oracle rate on new test files) further suggests that training data, prompt design, and tool integration influence testing discipline. While our data cannot isolate the
prompts behind each PR, oracle strength also varies with task type -- a proxy for what the agent was asked to do: feature work yields strong oracles in only 18.2\% of patches, versus 25.6\% for bug fixes and 24.9\% for test-focused tasks. This suggests practitioners obtain more meaningful
oracles when verification is scoped as an explicit objective rather than
bundled into feature PRs, and that adopting oracle-aware CI checks that
flag newly added test files lacking assertion patterns can surface this
gap early and separate verified contributions from structural scaffolding.

\section{Threats to Validity}
Our regex-based taxonomy classifies syntactic oracle signals in patch diffs, not complete source files; a modified-file patch classified as W1 may belong to a file with existing assertions, which we address by reporting newly created files separately. A patch classified as S1 confirms that an equality check exists, not that it checks the right property. The association between oracle strength and merge outcomes may be confounded by unobserved factors such as developer experience, merge-bot interactions, or CI requirements; we mitigate this through multivariate regression adjusting for agent, PR size, stars, task type, and language. Our dataset covers five agents active between December 2024 and July 2025, with OpenAI Codex contributing 65\% of PRs, and filters for repositories with 100+ stars; results may not generalize to smaller repositories or ecosystems with different testing conventions. 

Additionally, our taxonomy does not capture implicit oracles such as crash-on-failure or timeout-based detection, which some testing frameworks use as default verification mechanisms. Our sensitivity analysis shows that reclassifying W4 (mock/interaction) and W5 (snapshot) as strong shifts the overall weak rate by less than one percentage point, indicating that boundary decisions between categories do not materially affect the findings.

\section{Related Work}
Our work intersects three areas: the test oracle problem, LLM-based oracle generation, and empirical studies of agent-authored contributions. The oracle problem has a long research history \cite{barr2015oracle}. Coverage correlates only weakly with fault-detection effectiveness \cite{inozemtseva2014coverage}, and mature test suites exhibit gaps of up to 51 percentage points between executed and checked code \cite{hossain2023coverage}. Jain et al. \cite{jain2023oracle} formalized this as the ``oracle gap'', showing that high coverage alone does not capture test adequacy. Our study examines whether this distinction between executing code and verifying its behavior holds for agent-authored test code at scale. In the LLM setting, generated oracles frequently encode actual rather than expected behavior \cite{konstantinou2024llmoracles}, fine-tuned generators detect only a fraction of known faults \cite{fraser2025assert5}, and mutation-guided feedback improves quality but requires curated prompts \cite{khomh2024mutap}. These studies evaluate oracle quality in controlled settings; our work shifts the analysis to real-world agent-authored patches. For real-world agent behavior, Haque et al. measured test file presence on the AIDev dataset \cite{li2026aidev}, Milanese et al. compared test quality between agent and human PRs using test-smell detection, and Chowdhury et al. analyzed how code-review agents and feedback signal quality relate to PR outcomes \cite{haque2026tests, milanese2026human, chowdhury2026review}.

\section{Conclusion}

This study provides large-scale empirical evidence that test-file presence is an insufficient proxy for verification strength in agent-authored pull requests. By classifying 86,156 cumulative test-file patches across 2,807 repositories, we show that 80.2\% contain weak or no explicit oracle signals, with the strong-oracle rate varying from 18\% to 67\% across agents on newly created files. These findings suggest that coding agents generate test structure far more reliably than they generate oracle logic. Strong oracle signals coincide with larger, more heavily reviewed PRs; once this complexity is controlled, they predict higher merge likelihood, making oracle strength a positive signal masked by confounding rather than a merge disadvantage.

\bibliographystyle{IEEEtran}
\bibliography{references}
\end{document}